\documentclass[prd, amsfonts, twocolumn, nofootinbib, showpacs]{revtex4}
\usepackage{graphicx, epsfig}
\usepackage{color}
\usepackage{amsmath}
\usepackage{amssymb}
\newcommand{\be}{\begin{equation}}
\newcommand{\ee}{\end{equation}}
\newcommand{\bea}{\begin{eqnarray}}
\newcommand{\eea}{\end{eqnarray}}

\newcommand{\gapp}{\mathrel{\raise.3ex\hbox{$>$}\mkern-14mu
\lower0.6ex\hbox{$\sim$}}}
\newcommand{\lapp}{\mathrel{\raise.3ex\hbox{$<$}\mkern-14mu
\lower0.6ex\hbox{$\sim$}}}
\def\bbox{{\,\lower0.9pt\vbox{\hrule \hbox{\vrule height 0.2 cm
\hskip 0.2 cm \vrule  height 0.2 cm}\hrule}\,}}

\begin{document}
\title{Maximal temperature of the gas in AdS space-time}
\author{De-Chang Dai$^1$, Dejan Stojkovic$^2$}
\affiliation{ $^1$Institute of Natural Sciences, Shanghai Key Lab for Particle Physics and Cosmology, and Center for Astrophysics and Astronomy, Department of Physics and Astronomy, Shanghai Jiao Tong University, Shanghai 200240, China}
\affiliation{ $^2$ HEPCOS, Department of Physics, SUNY at Buffalo, Buffalo, NY 14260-1500}


\begin{abstract}
\widetext
Assuming only statistical mechanics and general relativity, we calculate the maximal temperature of gas of particles placed in AdS space-time.
If two particles with a given center of mass energy come close enough, according to classical gravity they will form a black hole. We focus only on the black holes with Hawking temperature lower than the environment, because they do not disappear. The number density of such black holes grows with the temperature in the system. At a certain finite temperature, the thermodynamical system will be dominated by black holes.  This critical temperature is lower than the Planck temperature for the values of the AdS vacuum energy density below the Planck density. This result might be interesting from the AdS/CFT correspondence point of view, since it is different from the Hawking-Page phase transition, and it is not immediately clear what effect dynamically limits the maximal temperature of the thermal state on the CFT side of the correspondence.
\end{abstract}


\pacs{}
\maketitle
In simple thermodynamics based on classical mechanics, there is no maximal temperature that a system cannot surpass. This fact does not change even with inclusion of special relativity. When gravity is included, it is usually believed that the maximal temperature that makes sense talking about is the Planck temperature \cite{sak}. However, this is an arbitrary cutoff and it does not follow from dynamics. Then in \cite{Dai:2016axz}, it was demonstrated that there exist a maximal achievable temperature in a system where particles obey the laws of (quantum) statistical mechanics and classical gravity. Namely, if two particles with a given center of mass energy come at the distance shorter than the Schwarzschild diameter apart, according to classical gravity they will form a black hole. It is possible to calculate that a simple thermodynamical system will be dominated by classical black holes at a critical temperature which is about three times lower than the Planck temperature. That represents the maximal achievable temperature in a simple thermodynamical system.

Calculations in \cite{Dai:2016axz} were performed for the gas in the flat space-time. The aim of this paper is to study a related question in anti-de Sitter (AdS) space-time. There are two crucial differences between the situations in flat and AdS space-times. First, the energy density of a gas of particles in AdS space depends on the energy spectrum of particles placed in the curved AdS background, which depends on the solution to the equation of motion in this background. Second, the metric for a black hole in an asymptotical ly AdS space is different from the metric in the flat space. Both of these properties crucially depend on the AdS radius. Thus, we expect the maximally achievable temperature to be AdS radius dependent too. In the following sections, we first discuss the thermodynamics of the gas, and the black hole solution in AdS space-time. We then give a condition for a production of a black hole in collision of two particles in an AdS background. We define the critical temperature at which the number density of created black holes (colder than the temperature of the gas) is greater than the number density of the particles. We finally numerically calculate the critical temperature as a function of the AdS radius.

We emphasis the difference between the effect we present here and the well-known Hawking-Page first order phase transition for thermal gravity in AdS space \cite{Hawking}. The Hawking-Page effect is a transition between the thermal gas and a single black hole placed in AdS space which, from the AdS/CFT correspondence point of view, can be interpreted as the confinement-deconfinement phase transition in the gauge theory \cite{Witten:1998zw}. Hawking and Page found the limiting temperature below which the black hole cannot exist and the system is dominated by the gas, and the temperature above which the black hole is strongly favored and the gas is prone to gravitational collapse. To calculate these temperatures, they considered equilibrium configurations between the gas and a single black hole in AdS space using global macroscopic quantities like entropy and free energy. In contrast, our black holes are produced in microscopic collisions of two particles. As a consequence, they could be produced even if they are not thermodynamically favorable. Our definition of the critical temperature is also different. Our system breaks down when we have more microscopic black holes than particles per unit volume, while the Hawking-Page critical temperature is dictated by the equilibrium between the gas and a single large black hole. Finally, our effect is present even in flat space-time \cite{Dai:2016axz}, while the Hawking-Page effect crucially relies on the natural box that AdS space provides and allows for the black hole - gas equilibrium.
Thus, in this paper we are describing related, but substantially different phenomenon.

\section{Particle energy in AdS space}
The metric of the anti de-Sitter (AdS) space can be written in a static form as
\begin{eqnarray} \label{metric}
&&ds^2=-V(r)dt^2+V(r)^{-1}dr^2+r^2d\Omega^2\\
&&V(r)=(1+\frac{r^2}{b^2}) \label{pot}\\
&&d\Omega^2=d\theta^2+\sin^2\theta d\phi^2\\
&&b=\Big(-\frac{3}{\Lambda}\Big)^{1/2}
\end{eqnarray}
$\Lambda$ is the absolute value of the cosmological constant (or vacuum energy density) in the AdS space. All the relevant quantities are expressed in units of Planck mass, $m_{pl}$. One peculiar feature of AdS space is that the gravitational potential relative to any origin increases with the distance from the origin, as can be seen from Eq.~(\ref{pot}).

A thermal state in AdS space can be constructed by periodically identifying the imaginary time coordinate with period $\beta$. The corresponding temperature is then $T=\beta^{-1}$ \cite{Hawking}. The locally measured temperature at any given point $r$ is
\begin{equation}\label{lt}
T_{r}=\frac{T}{\sqrt{V(r)}}  .
\end{equation}
This means that the local temperature is red-shifted by the gravitational potential and decreases
like $1/r$ for $r \gg b$. Thus we expect that the energy of massless particles will fall of as $1/r^4$, and the total energy of the thermal state will be finite.
The fact that the locally measured temperature (and energy) gets infinitely redshifted as $r \rightarrow \infty$ can be interpreted as an infinite potential wall that AdS space has at asymptotic infinity.

For our purpose, we need to calculate the energy distribution function for particles in AdS space. For simplicity, we consider a conformally coupled massless scalar filed.
According to \cite{Hawking}, we can take a thermal state from the AdS space to the Einstein's static universe and perform calculations there.
The AdS metric (\ref{metric}) can be transformed into
\begin{equation}
ds^2=\frac{1}{\cos^2\chi}\Big(-d\tau^2 +b^2(d\chi^2+\sin^2\chi d\Omega) \Big) .
\end{equation}
We omit the cumbersome expressions for coordinate transforms, which can be found in many textbooks anyway.
The metric inside the parentheses represents the Einstein universe with the fixed radius $b$.
\begin{equation} \label{es}
d\tilde{s}^2=-d\tau^2+b^2(d\chi^2 +\sin^2\chi d\Omega^2).
\end{equation}
This metric has the same thermal states as the original AdS space, since the theory of conformally coupled massless particles is conformally invariant.

To determine the energy density of gas of particles in Einstein's universe, we have to find a solution for the particle's wave function in the background of Eq.~(\ref{es}). Several different possibilities have been studied in \cite{Shrodinger,Altaie:1978dx,Altaie:2002tv,Altaie:2003ef}. In our case, we have a massless scalar field, $\phi$, which satisfies the conformally invariant covariant Klein-Gordon equation
\begin{equation}
\phi^{;\alpha}{}_{;\alpha}-\frac{1}{6}R\phi=0  ,
\end{equation}
where $R$ is the scalar curvature of the metric (\ref{es}). The particle energy eigenvalues \cite{Altaie:1978dx} are
\begin{equation}
\epsilon_N =\frac{N+1}{b}\text{, $N=0,1,2...$}
\end{equation}
where the discrete index $N$ labels the energy eigenvalues.
 The degeneracy number is given by $d_N=(N+1)^2$.

For massless bosons, the energy distribution function at a given temperature, $T$, is given by
\begin{equation}
F(E,T)=\frac{1}{e^{\frac{E}{T} }-1} .
\end{equation}
The energy density of massless scalar particles in Einstein's universe is therefore
\begin{equation}
\rho(r=0)=\frac{1}{2\pi^2 b^3}\sum_{N=0}^\infty \epsilon_N (N+1)^2F(\epsilon_N,T) .
\end{equation}

To obtain the corresponding energy distribution in AdS space, we have to account for the redshift effect which depends on the location, $r$, of the particle in AdS space. After including the redshift, the local energy of a scalar particle in AdS space is
\begin{equation} \label{rs}
\epsilon_N(r)=\frac{\epsilon_N}{\sqrt{1+\frac{r^2}{b^2}}}
\end{equation}
where the redshift factor is just the $g_{tt}$ component of the AdS metric in Eq.~(\ref{metric}).
The local energy density is
\begin{equation}
\rho(r)=\frac{1}{2\pi^2 b^3 (1+\frac{r^2}{b^2})^2 }\sum_{N=0}^\infty \epsilon_N (N+1)^2F(\epsilon_N(r),T_r) ,
\end{equation}
where the expression for the local temperature, $T_r$, is defined in Eq.~(\ref{lt}).
The local number density of particles with energy $\epsilon_N(r)$ is
\begin{equation}
n(r,\epsilon_N)=\frac{1}{2\pi^2 b^3(1+\frac{r^2}{b^2})^{3/2}} (N+1)^2F(\epsilon_N,T_r)
\end{equation}

\begin{figure}
   \centering
\includegraphics[width=8cm]{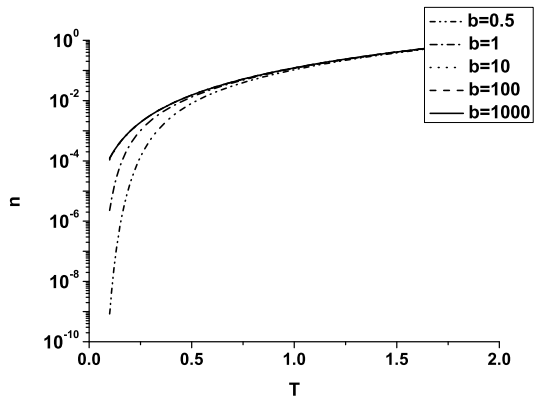}
\caption{The particle number density, $n$, in AdS space as a function of the temperature, $T$, is plotted for several different values of AdS radius, $b$. The three curves for $b=1000$, $b=100$ and $b=10$  almost overlap with the flat space case. This is expected since since the vacuum energy density of AdS space, $\Lambda$, is inversely proportional to $b$. The number densities corresponding to lower values of $b$, (i.e. $b=1$ or $b=0.5$), are also lower than the number density in the flat space for temperatures $T\lessapprox b^{-1}$, but they become very close to the flat space value for $T\gtrapprox b^{-1}$. The number density is given in units of $m_{pl}^{-3}$.}
\label{number-density}
\end{figure}

Note that these quantities are functions of location $r$, since the metric (\ref{metric}) is $r$-dependent.
For illustration, we plot the particle number density as a function of the temperature, $T$, at $r=0$ in Fig.~\ref{number-density}.  For temperatures higher than the inverse AdS radius, i.e. $T \gg b^{-1}$, the number density reduces to the flat space case. For example, we can see that the number density curves corresponding to $b=1000$, $b=100$, and $b=10$ are indistinguishable from the flat space case. This is expected since that case corresponds to the temperatures much higher than the vacuum energy density of AdS space, $\Lambda$. The difference is more pronounced for lower values of $b$, i.e. higher values of $\Lambda$, while the general behavior stays the same. For example, the number densities corresponding to $b=1$ or $b=0.5$, are lower than the number density in the flat space for temperatures $T\lessapprox b^{-1}$, while they become very close to the number density in the flat space for $T\gtrapprox b^{-1}$. The reason for this behavior is that the AdS space is compact, as opposed to the flat space.

\section{AdS-Schwarzschild black holes}
We now study black holes placed in AdS space.
The metric of a static black hole of mass $M$, in a space which is asymptotically AdS with the AdS radius $b$, is given by
\begin{eqnarray}
&&ds^2=-A(r)dt^2+A(r)^{-1}dr^2+r^2d\Omega^2\\
&&A(r)=(1-\frac{2M}{r}+\frac{r^2}{b^2})
\end{eqnarray}
The black hole gravitational radius can be found by solving the equation $A(r)=0$. This gives
\begin{equation}
r_+(M)=\frac{b^{\frac{2}{3}}}{3^{\frac{2}{3}}}\frac{(9M+\sqrt{81M^2+3b^2})^{\frac{2}{3}}-3^{\frac{1}{3}}b^{\frac{2}{3}}}{(9M+\sqrt{81M^2+3b^2})^{\frac{1}{3}}} .
\end{equation}
The temperature of this black hole is given as
\begin{equation} \label{tbh}
T_b^{-1}=\frac{4\pi b^2 r_+}{3r_+^2 +b^2} .
\end{equation}

\begin{figure}
   \centering
\includegraphics[width=8cm]{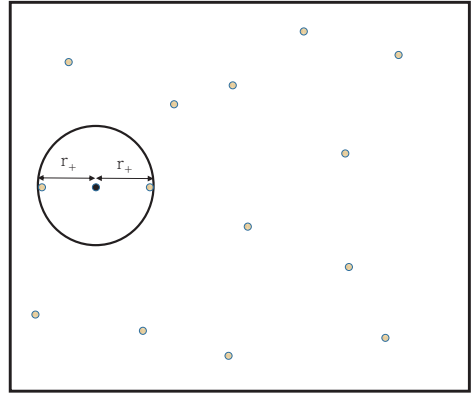}
\caption{In the flat space-tme, if two particles with the center of mass energy $M$ come to a distance $r$ which is shorter than $2r_+(M)$ ($r_+$ is the gravitational radius for the given mass), then according to the hoop conjecture a black hole will be formed.}
\label{collision}
\end{figure}

In flat space, if two particles with the center of mass energy $M$ come to a distance $r$ which is shorter than twice the gravitational radius for the given mass (like in Fig.~\ref{collision}), then according to the hoop conjecture a black hole will be formed \cite{hoop,Dai:2007ki}. This however must be modified in AdS space, because of the existence of the cosmological constant which changes the geometry of the space-time.

\begin{figure}
   \centering
\includegraphics[width=7cm]{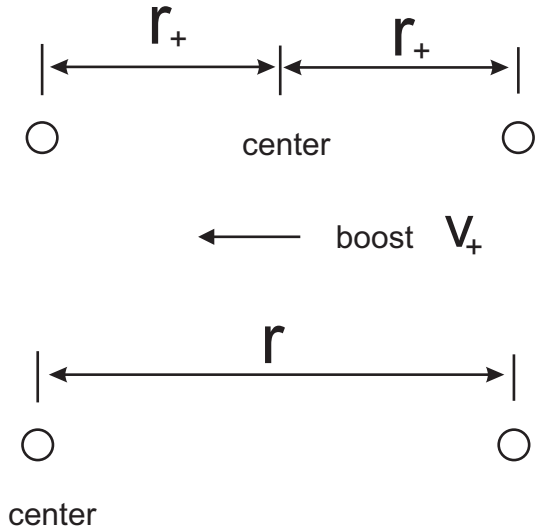}
\caption{Collision of two particles in AdS space-time which creates a black hole located at the coordinate center. The gravitational radius of the AdS black hole is $r_+$ (upper part of the figure). To find the coordinate distance, $r$, between these two particles which will result in the black hole formation, we have to take into account that the particle energy as seen by an observer located at the center appears blueshifted due to the presence of the cosmological constant which modifies the geometry. This results in somewhat larger  distance which yields the black hole formation than naively expected.}
\label{boost}
\end{figure}

Consider a black hole with the gravitational radius $r_+$ which is created in AdS space at the coordinate center $r=0$, as shown in the upper part of the Fig.~\ref{boost}. In that system, the particles that produced this black hole were located at most at $\pm \vec{r}_+$ (at two opposite sides from the origin). To simplify the calculations, we now go to the referent system where one of the colliding particles is located at $r=0$. In flat space-time, the position of the second particle which results in the black hole formation would be simply $2 \vec{r}_+$, like in Fig.~\ref{collision}. However, in curved space-time we cannot simply add two vectors, so we have to solve the following problem. During the collision that produced the black hole, one of the particles was located at the coordinate center, $r=0$, while the other one was at the coordinate distance $r$ (lower part of the Fig.~\ref{boost}). An observer located at the center sees that the energy of the particle at $r$ is blueshifted by a factor
\begin{equation}
\gamma=\sqrt{1+\frac{r^2}{b^2}} \, ,
\end{equation}
which follows straight from the AdS metric in Eq.~(\ref{metric}).
This is equivalent to the second particle moving toward the center with velocity, $v$, which satisfies
\begin{equation}
\label{r1}
\frac{1}{\sqrt{1-v^2}}=\sqrt{1+\frac{r^2}{b^2}} \, .
\end{equation}
Accordingly, the particle located at $r_+$ has the corresponding velocity, $v_+$, which satisfies
\begin{equation}
\label{r2}
\frac{1}{\sqrt{1-v_+^2}}=\sqrt{1+\frac{r_+^2}{b^2}} \, .
\end{equation}
Since in this process we boosted the system in which the black hole is in the center by velocity $v_+$, then $v$ and $v_+$ must satisfy
\begin{equation}
\label{r3}
v=\frac{2v_+}{1+v_+^2}  \, ,
\end{equation}
which is just the relativistic addition of velocities.
From equations (\ref{r1}), (\ref{r2}) and (\ref{r3}), one finds
\begin{equation}
r=2r_+\sqrt{1+\frac{r_+^2}{b^2}}
\end{equation}

This implies that if a particle is located at the center of the AdS space, the other particle must be within the coordinate distance $r$, where
\begin{equation}
r<R(M)=2r_+\sqrt{1+\Big(\frac{r_+}{b}\Big)^2},
\end{equation}
in order to form a black hole. This value is always larger than the flat space value, even for black holes with $r_+ < b$ (most of the black holes produced in collisions of particles will be in that regime). As a consequence, the probability for two particles to form a black hole in collisions in AdS space will be larger than that in flat space.

\section{Black hole domination over particles and maximal temperature in AdS}

The number of particles, $\Delta N_p$, in the region of space labeled by  $(\Delta r, \Delta \theta, \Delta \phi)$, is
\begin{equation}
 \Delta N_p (\epsilon_N) =n(r,\epsilon)r^2\sin\theta\Delta r \Delta \theta \Delta \phi .
\end{equation}

We now single out two particles which can interact and make a black hole.  A particle with energy and momentum $(E_1, \vec{k}_1)$ located at the origin can interact with a particle with energy and momentum $(E_2,\vec{k}_2)$ located at some distant point $r$. Their center of mass energy is
 \begin{eqnarray}
 M&=&\sqrt{2E_1E_2 -2\vec{k}_1\cdot \vec{k}_2}\\
 \label{mass}
 &=&\sqrt{2E_1E_2 -2E_1E_2\cos\alpha} ,
\end{eqnarray}
where $\alpha$ is the angle between $\vec{k}_1$ and $\vec{k}_2$. We assume $\alpha$ is isotropic in the solid angle distribution.
The black hole number density is therefore
\begin{equation} \label{nbh}
\left. n_{bh} = \frac{1}{2}\int \sum_{N_1,N_2}n(0,\epsilon_{N_1}) n(r,\epsilon_{N_2})4\pi r^2  dr \frac{\sin\alpha d\alpha}{2} \right|_{r<R(M)}
\end{equation}
The black hole mass, $M$, is defined in Eq.~(\ref{mass}). The term $\frac{\sin\alpha d\alpha}{2}$ represents the flat distribution of the angle $\alpha$.
The summation in discrete energy eigenvalues $N_1$ and $N_2$ is performed numerically up to $E/T<15$, which gives an accuracy of $e^{-15}$.
Black holes which are hotter than the background temperature, i.e. $T_b > T$, will evaporate and disappear. Therefore,
we count only black holes with temperature $T_b<T$, where the temperature of the black hole is given by Eq.~(\ref{tbh}). For such black holes accretion of particles dominates and they will not disappear. Therefore, the numerical code that integrates Eq.~(\ref{nbh}) checks that a black hole has temperature lower than the gas, i.e. $T_b<T$, and if this is not true, the code discards that black hole.

\begin{figure}
   \centering
\includegraphics[width=8cm]{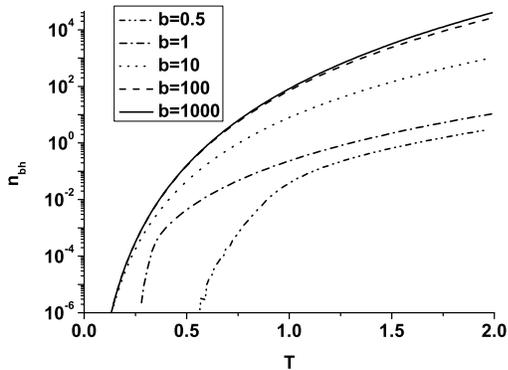}
\caption{For large values of $b$ (e.g. $b=1000$, solid line) the created black hole number density, $n_{bh}$, is the same as in the flat space. For lower values of $b$, the creation of black holes is suppressed, so $n_{bh}$ is lower than in the flat space. Even at high temperatures, low values of $b$ still affect the black hole density because of the energy redshift effect in Eq.~(\ref{rs}).}
\label{bh-number-density}
\end{figure}

The plots for the created black holes number density, $n_{bh}$, are shown in Fig.~\ref{bh-number-density}. For large values of the AdS radius $b$ (i.e. small values of the cosmological constant), $n_{bh}$ is the same as in the flat space. For lower values of $b$, the creation of black holes is suppressed, so $n_{bh}$ is lower than in the flat space. We can see that even at high temperatures, the effect of low $b$ still affects the black hole production. This is because the energy redshift effect in Eq.~(\ref{rs}) is stronger for lower values of $b$. For large distances, $r$, the black hole production suppression is also stronger.

We are now ready to extract the maximal temperature of the gas in AdS space. We can define a critical temperature $T_c$ at which
the black hole number density, $n_{bh}$, becomes higher than the particle number density, $n$. Beyond that point one cannot heat up the gas anymore since the space is dominated by black holes.

We note that this is a conservative approach, which yields a conservative upper limit on the temperature. At any given temperature of the gas, we count only those black holes whose Hawking temperature is lower than the environment. Such black holes will not evaporate and will remain in the system. We do not take accretion into account, but this is a conservative treatment, since a black hole is more likely to accrete particles than other black holes as long as there are more particles than black holes in the system. Above the crossover temperature, when there are more black holes, a single black hole is more likely to accrete (merge with) other black holes, but at that time the system is already dominated by black holes so redistribution of energy between the individual black holes will not reverse domination. Also, at that point, pumping more energy into the system goes primarily into black holes not particles.
Therefore, our estimate will yield a conservative upper limit on the temperature.

\begin{figure}
   \centering
\includegraphics[width=8cm]{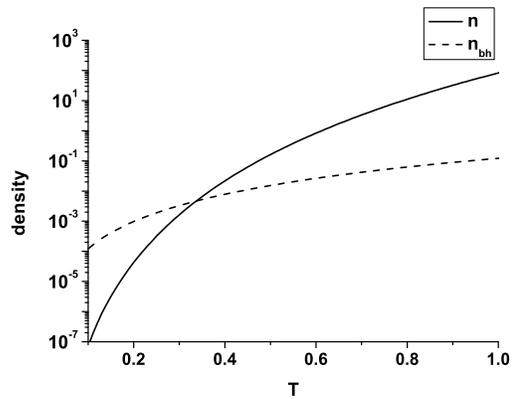}
\caption{We can define a critical temperature $T_c$ at which
the black hole number density, $n_{bh}$, becomes higher than the particle number density, $n$. At low temperatures, the black hole density is lower than particle number density. At high temperatures this is reversed. The crossover defines the critical temperature.  If we set $b=1000$, we find the critical temperature $T_c\approx 0.34 m_{pl}$.}
\label{compare}
\end{figure}

\begin{figure}
   \centering
\includegraphics[width=8cm]{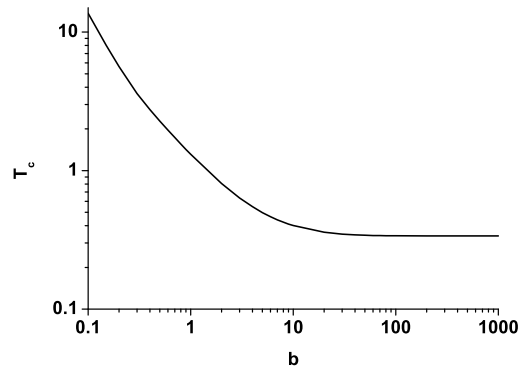}
\caption{The critical temperature, $T_c$, as a function of AdS radius $b$. Obviously, $T_c$ is increasing as the value of $b$ is decreasing. This means that larger values of the AdS vacuum energy density allows higher maximal temperature of the gas. }
\label{cross}
\end{figure}

We show the crossover between the particles and black holes domination in Fig.~\ref{compare}. For example, if we set $b=1000$, we find the critical temperature $T_c\approx 0.34 m_{pl}$.
It is also interesting to see how the critical temperature changes with $b$. This trend is shown in Fig.~\ref{cross}. Clearly, the critical temperature, $T_c$, is increasing as the value of $b$ is decreasing. So large values of the AdS vacuum energy density alow higher maximal temperature of the gas.

As it can be seen in Fig.~\ref{cross}, for values of $b$ larger than approximately one (in Planck units), the critical temperature is lower than the Planck temperature, and asymptotically matches the value in flat space-time \cite{Dai:2016axz}. In the opposite case where $b$ is larger than one in Planck units (which implies the vacuum energy density of AdS space higher than the Planck energy density), the critical temperature goes above the Planck value. This is however regime where quantum gravity effects become very important and we do not have an explicit control over our calculations.

\section{Conclusions}

In this paper we investigated the question of a maximal temperature of gas of particles that AdS space-time can allow.
In the context of string theory, it is well known that at very high temperatures the density of states grows exponentially indicating a phase transition at which very long strings are copiously produced \cite{Hagedorn:1965st,Atick:1988si,Alvarez:1985fw,Bowick:1985az}. This Hagedron temperature (which explicitly depends on the stringy parameters in the model) could represent a maximal achievable temperature because any increase in energy of the system would go into creating new stringy states rather than increasing the temperature (for an alternative point of view see \cite{Dienes:2005vw}).
In our approach, however, we do not rely on string theory, we assume only statistical mechanics and general relativity.
At any given non-zero temperature, there is a finite probability that two particles will collide and form a mini black hole. This probability grows as the temperature increases, which in turn can have interesting consequences (see e.g. \cite{Dai:2016axz,Dong:2016kuv}). In our context, there is a critical temperature where the number density of the created black holes (colder than the gas) becomes higher than the number density of particles in the gas. This critical temperature represents a maximal achievable temperature in AdS space-time. We show that this temperature crucially depends on the AdS radius $b$. For large values of $b$ (e.g. $b\gtrapprox 1$ in the Planck units), the critical temperature is lower than the Planck temperature, while for small values (e.g. $b\lessapprox 1$ in the Planck units, which implies the vacuum energy density of AdS space higher than the Planck energy density), the critical temperature goes above the Planck value. In that regime, however, quantum gravity effects become very important, and in the absence of the full theory of quantum gravity we do not know what happens there. Of course, if some new physics enters below the Planck scale, one would have to do the calculations in that new framework. In the concrete case of the string theory, one would have to calculate the probability of black holes (or string balls or other non-perturbative states) creation by highly excited string states.

This result might be interesting from the AdS/CFT correspondence point of view. On the CFT side, there is no gravity and black holes are not included. It would be therefore interesting to see what effect dynamically limits the maximal temperature of the thermal state on the CFT side of the correspondence (perhaps creation of some non-perturbative states?). This is different from the Hawking-Page effect, which is a transition between the thermal gas and a single black hole placed in AdS space. That effect can be interpreted from the AdS/CFT correspondence point of view as the confinement-deconfinement phase transition in the gauge theory. However, the Hawking-Page effect follows from examinations of  the global macroscopic quantities like entropy and free energy. In contrast, our black holes are produced in microscopic collisions of two particles. As a consequence, they could be produced even if they are not thermodynamically favorable. Our definition of the critical temperature is also different. Our system breaks down when we have more microscopic black holes than particles per unit volume, while the Hawking-Page critical temperature is dictated by the equilibrium between the gas and a single large black hole. In a sense, the Hawking-Page effect is kinematic in its nature, while ours is dynamical. Thus, these two effects might be somewhat related, but the details are certainly different.

From Eq.~(3.3) in \cite{Hawking} we see that the Hawking-Page phase transition happens at $T \sim 1/\sqrt{b}$. Above that temperature the gas cannot support itself and eventually collapses into a single large black hole. From our Fig.~\ref{cross} we see that  $T \sim 1/\sqrt{b}$ is lower than our critical temperature, $T_c$, (at least for $b\gtrapprox 1$), which implies that our effect happens after the Hawking-Page transition (going from low to high temperatures).  From the geometric point of view, the formation of a large black hole in gravitational collapse of some distribution of matter in AdS space is no different from formation of an apparent horizon in any other isotropic universe. An outside observer will see an AdS-Schwarzschild black hole, but an observer inside the horizon will witness all the time evolution of the gravitational collapse in which the collapsing gas goes through stages of increasing densities and temperature. From the analysis we performed here, it appears that our effect is mostly relevant for an observer inside the horizon.

\begin{acknowledgments}
D.C Dai was supported by the National Science Foundation of China (Grant No. 11433001 and 11447601), National Basic Research Program of China (973 Program 2015CB857001), No.14ZR1423200 from the Office of Science and Technology in Shanghai Municipal Government and the key laboratory grant from the Office of Science and Technology in Shanghai Municipal Government (No. 11DZ2260700) and  the Program of Shanghai Academic/Technology Research Leader under Grant No. 16XD1401600. D.S. was partially supported by the US National Science Foundation, under Grant No. PHY-1417317.
\end{acknowledgments}

\end{document}